\documentclass[aps,pra,floatfix,showpacs,noshowkeys,epsfig,graphics,natbib]{revtex4-1}
\usepackage{graphicx}
\usepackage{amsmath}
\usepackage{amsfonts}
\usepackage{amssymb}
\usepackage{amsthm}

\newcommand{\newc}{\newcommand}
\newc{\beq}    {\begin{equation}}
\newc{\eeq}    {\end{equation}}

\newc{\beqa}    {\begin{eqnarray}}
\newc{\eeqa}    {\end{eqnarray}}
\newc{\bs}    {\section}
\newc{\no}    {\\ \nonumber}

\newtheorem{theorem}{Theorem}

\newc{\st}    {\stackrel}

\begin{document}
\title{ On the origin of the holographic principle}
\author{Jae-Weon Lee}\email{scikid@jwu.ac.kr}
\affiliation{ Department of energy resources development,
Jungwon
 University,  5 dongburi, Goesan-eup, Goesan-gun Chungbuk Korea
367-805}

\date{\today}

\begin{abstract}
It was recently  suggested
that  quantum mechanics and gravity
are  not fundamental but emerge from information loss at causal horizons.
On the basis of the formalism
the holographic principle is also shown to arise naturally from
the loss of information about bulk fields observed by
an outside observer.
As an application, Witten's prescription is derived.

 \end{abstract}

\maketitle
%\section{Introduction}

\section{introduction}

The holographic principle~\cite{hooft-1993,susskind-1995-36}, including the AdS/CFT correspondence~\cite{Maldacena},
asserts an unexpected connection between the physics in a bulk
and quantum field theory (QFT) on its boundary surface. The origin of this mysterious connection is  still unknown.
On the other hand, many studies of black hole physics after  Bekenstein and Hawking  have
 implied a deep connection between gravity and  thermodynamics~\cite{Bekenstein}.
For example, Jacobson suggested that Einstein's equation describes thermodynamics at
   local Rindler horizons, and Padmanabhan ~\cite{Padmanabhan:2009kr}   proposed that
 classical gravity can be derived from the equipartition energy of horizons.
 Verlinde recently  proposed an intriguing
 idea~\cite{Verlinde:2010hp} linking both Einstein's gravity and Newton's mechanics to entropic forces.
All these works emphasized the strange connection between thermodynamics and gravity
the origin of which is also still mysterious
~\cite{Zhao:2010qw,Cai:2010sz,Cai:2010hk,Myung:2010jv,Liu:2010na,Tian:2010uy,Pesci:2010un,Diego:2010ju,
Vancea:2010vf,Konoplya:2010ak,Culetu:2010ua,Zhao:2010vt,Ghosh:2010hz,Munkhammar:2010rg,Kuang:2010gs,Banerjee:2010yd}.
Since the thermodynamic entropy can be interpreted as a measure of information or information loss, the
connection
implies a close relationship between information and gravity.

In a series of works ~\cite{myDE,Kim:2007vx,Kim:2008re,Lee:2010bg}, based on
information theory, my colleagues and I suggested
that  information loss (or  quantum entanglement) at causal horizons is the key to understanding the origins of dark energy~\cite{myDE}, black hole mass~\cite{Kim:2007vx}
and even  Einstein's gravity~\cite{Lee:2010bg}.
Many other studies in quantum information science have supported the idea that information is fundamental.
For example, one can obtain a quantum mechanical state discrimination bound
from the condition that information propagation is not superluminal (see, for example, ~\cite{PhysRevA.78.022335}).
Landauer's principle in quantum information science implies that information is  physical~\cite{landauer}.
 Zeilinger and Brukner ~\cite{zeilinger1999,brukner-1999-83} suggested that quantum randomness arises from the discreteness of information. 't Hooft also suggested that quantum mechanics has a deterministic theory
 that includes local information loss~\cite{hooft-2002}.

Since
the number of degrees of freedom (DOF) in a bulk theory
is proportional to the volume of the bulk, whereas the number of  DOF
in a boundary theory is proportional to its surface area,
it seems impossible to  prove a generic holographic principle within the context of QFT.
It might be possible to prove the principle on the basis of
 new principles that can explain the origin of both gravity
and quantum mechanics.

In \cite{myquantum} I showed that  quantum mechanics
is not fundamental but emerges from  information loss at  causal horizons.
If gravity and quantum  mechanics can be derived by considering information loss
at causal horizons (see \cite{myreview} for a review), it is natural
 to think that the holographic principle has a similar origin.
%Thus, information loss could be a key to proving the holographic principle.
This paper  suggests that the  principle
can be derived by applying information theory
to causal horizons.

 In Sec. II, the derivation of  quantum mechanics from information theory is reviewed.
In Sec. III, a derivation of  the holographic principle  is
presented, and Witten's prescription is derived as an application.
Sec. IV contains conclusions.

\section{Quantum field theory from information loss}

In this section, I review the way in which  QFT, and hence
quantum mechanics,  arises from the application of
 information theory at causal horizons.
It is important to understand that our theory is based on neither   QFT nor  Einstein's gravity.
On the contrary, they could be derived from the postulates below.

Inspired by the works described in Sec. I, we can choose the following postulates
as new
general guiding principles on which any physical law should be based.\\
\begin{enumerate}
{\it
\item  General equivalence:
All systems of reference (coordinates) are equivalent for formulating physical laws regardless of their motions.
\item Information  has a finite density and speed; the quantity of information contained in a finite object is finite, and there is a maximal speed of classical information propagation, namely, the light velocity.
\item  Information is fundamental:
 Physical laws regarding an object (matter or spacetime) should be such that  they
 respect  observers' information about the object.
 }
\end{enumerate}

We also assume the metric nature of spacetime (however, we do not assume Einstein's equation)
and ignore any fluctuation of spacetime in this paper.
Postulate 2 implies that  for a given observer, there could be causal horizons
that block information
about a region the observer cannot access.

Although  postulates 1 and 2 are familiar, postulate 3 deserves more explanation.
It implies that interpretation of a physical reality depends on the information an observer can access,
and hence there is no objective reality independent of observers. This sounds counterintuitive, but
it is exactly what ordinary quantum mechanics says. For example, it is possible that a pure qubit state
$\psi=(|0\rangle +|1\rangle)/\sqrt{2}$ seen by one observer
 can be a maximally mixed state  for another observer who could not access  information about the state.
An important point here is that both descriptions of the same qubit are perfectly valid and not in contradiction.
Similarly, regardless of measurements done by an observer inside a causal horizon,
quantum states of matter inside the horizon seen by
an outside observer are maximally mixed. Two descriptions of the matter including the observer's state itself
should be coincident.
 On average, no observer has priority.
Considering the surface action terms in relativity,  Padmanabhan  also  pointed out that   physical theories must be formulated
 in terms of variables any given observer can access~\cite{Padmanabhan:2003ub}.

One special conclusion derived from the three postulates together
 is that physics inside a causal horizon
should respect (or be consistent with) the ignorance of an outside observer about the inside region. This naturally introduces the notion
of a horizon entropy arising by definition from information loss.
Some authors have argued that this information loss is the origin of black hole entropy
~\cite{PhysRevD.34.373}.

Using the above postulates, I showed in  \cite{myquantum}
that quantum mechanics is not fundamental but emerges from
 the application of classical information theory to causal horizons.
The path integral (PI) quantization and quantum randomness can be derived by considering information
loss for accelerating observers
of fields or particles crossing Rindler horizons. This implies that information
 is one of the fundamental roots of all physical phenomena.
  I also investigated the connection between this theory and Verlinde's entropic gravity theory.

Let us briefly review the information theoretic formalism suggested in Ref. \cite{myquantum}.
Consider an
accelerating observer $\Theta_R$ with acceleration $a$ in the $x_1$ direction
 in a flat spacetime with coordinates $x=(t,x_1,x_2,x_3)$ (See Fig. 1).
The Rindler coordinates $\xi=(\eta,r,x_2,x_3)$ for the observer are
\beq
\label{rindler}
t= r~ sinh (a \eta),~ x_1= r ~cosh (a \eta).
\eeq
There is another observer $\Theta_M$ inside the Rindler horizon.
Now, consider a field $\phi$
 crossing the Rindler horizon
  and entering the future wedge $F$. The configuration for
 $\phi(x)$ in $F$ is just a scalar function of the coordinates $x$, not a classical field.

\begin{figure}[tpbh]
\includegraphics[width=0.3\textwidth]{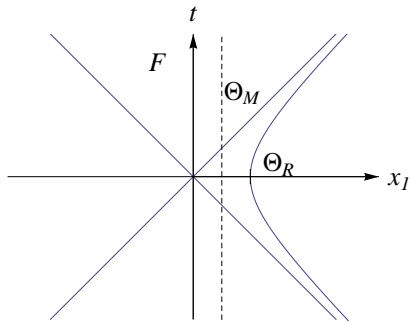}
\caption{ A Rindler  observer $\Theta_R$ (curved line) has no accessible information about the  field $\phi$
in  a causally disconnected  region $F$.
The observer can only estimate a probabilistic distribution of the field,
which turns out to be equal to that of a quantum field seen by a Minkowski observer $\Theta_M$ (dashed line).
}
\end{figure}

As the field enters  the Rindler horizon for the observer $\Theta_R$,
 the observer receives no further information about  future configurations  of $\phi$.
All that the observer can guess about the evolution of $\phi$
 is the probabilistic distribution $P[\phi]$ of $\phi$ beyond the horizon.
The information already known about $\phi$  constrains $P[\phi]$.
According to our postulates,  the physics
in the wedge $F$ is determined
not by deterministic classical physics  but by the evolution of information.
 The maximum ignorance about the field can be mathematically expressed by maximizing the Shannon
information entropy
$
h[P]=-\sum_{i=1}^n P[\phi_i] ln  P[\phi_i]
$ of the possible (discrete) configurations
$\{\phi_i(x)\}~, i=1\cdots n$ that the field may take with a probability $P[\phi_i]$.
 If there is  information for $\Theta_R$ represented by
 $N_j$ testable expectations
\beq
\label{constraint}
\langle f_j\rangle\equiv  \sum_{i=1}^n P[\phi_i] f_j[\phi_i],~(j=1\cdots N_j),
\eeq
 we should use Boltzmann's theorem of maximum entropy to calculate the probability distribution
 $P[\phi]$.
Here, $f_j,~(k=1\cdots N_j)$ is a functional of $\phi$  and $\langle f_j\rangle$ is its statistical expectation value
with respect to $P[\phi]$. According to the theorem,
 by maximizing the Shannon entropy
with the constraints in Eq. (\ref{constraint}),
one can obtain
 the following probability distribution
 \beq
 \label{P}
 P[\phi_i]=\frac{1}{Z} exp\left[-\sum_{j=1}^{N_j} \lambda_j f_j(\phi_i)\right]
 \eeq
with a partition function
$
Z=\sum_{i=1}^n  exp\left[-\sum_{j=1}^{N_j}\lambda_j f_j(\phi_i)\right].
$

One constraint may come from the energy conservation
%\beq
$\sum_{i=1}^n  P[\phi_i]H(\phi_i) = E,$
%\eeq
where $H(\phi_i)$ is the Hamiltonian as a function of the field configuration $\phi_i$ and $E$ is its expectation,
%This comes from the fact that the energy expectation value of the field should not change.
  and another trivial one is the unity of  the probabilities
$
\sum_{i=1}^n  P[\phi_i] = 1.
$
Then, the probability with the constraints
 estimated by the Rindler observer should be
\beq
\label{P}
P[\phi_i] = \frac{1}{Z} \exp\left[- \beta H(\phi_i)\right],
\eeq
where $\beta$ is a Lagrangian multiplier, and
 the partition function is
%\beq
%\label{Z}
 $Z = \sum_{i=1}^n  \exp\left[- {\beta H(\phi_i)} \right]=tr~ e^{-\beta H}$.
 %\eeq%
 Thus, the thermal nature of quantum fields is a natural consequence of classical information theory, when
 information loss with constraints occurs.

 As an example, let us  consider  a scalar field with Hamiltonian
 \beq
 H(\phi)=\int d^3 x \left[ \frac{1}{2}\left( \frac{\partial \phi}{\partial t}\right)^2+
 \frac{1}{2}\left( {\nabla \phi}\right)^2 + V(\phi) \right]
 \eeq
 and a potential $V$.
 %$H$ alone, without a guiding principle, does not fully give us dynamics, neither classical nor quantum.
  For the Rindler observer with the coordinates $(\eta,r,x_2,x_3)$  the proper time variance is
  $ard\eta$ and the Hamiltonian becomes
  \beq
 H_R =  \int_{r>0} dr dx_\bot~ ar
 [
  \frac{1}{2}\left( \frac{\partial \phi}{ar \partial \eta}\right)^2
 +
 \frac{1}{2}
 \left(\frac{\partial \phi}{ \partial r} \right)^2+\frac{1}{2}
 \left( {\nabla_\bot \phi}\right)^2
 +  V(\phi)] ,
 \eeq
where $\bot$ denotes the plane orthogonal to $(\eta,r)$ plane.
Then, $Z$ becomes Eq. (2.5) of Ref. ~\cite{1984PhRvD..29.1656U};
\beq
\label{Z_R}
 Z_R = tr~ e^{-\beta H_R}.
 \eeq
Notice
  that $Z$ (and hence $Z_R$) here is not a quantum partition function but a statistical one
   corresponding to the uncertain
 field configurations  beyond the horizon.

The equivalence of this form of  $Z_R$ and  a quantum partition function for a  scalar field
in the Minkowski spacetime (say $Z_Q$)
 is  shown in Ref. ~\cite{1984PhRvD..29.1656U}, which is the famous Unruh effect.
(See ~\cite{Crispino:2007eb}  for a review.)
A continuous version of Eq. (\ref{Z_R}) in QFT is
 \beq
 Z_R =\no
    N_0\int_{\phi(0)=\phi(\beta)}
  D\phi ~exp\left\{-\int_0^{\beta} d\tilde{\eta} \int_{r>0} dr dx_\bot~ ar H_R \right\}.
 \eeq
  By further  changing integration variables as
$\tilde{r}=r~ cos(a\tilde{\eta}),  \tilde{t}= r~sin(a\tilde{\eta})$ and choosing $\beta=2\pi/a$
the  region of integration is transformed
into the full two dimensional flat space, which leads to Unruh temperature
$ T_U={ a\hbar}/{2 \pi k_B }$, where $k_B$ is the Boltzman constant.
Then, the partition function becomes that of an euclidean flat spactime;
\beqa
\label{ZEQ}
 Z^E_Q &=&  N_1\int
  D\phi ~exp\{- \int d\tilde{r}d\tilde{t} dx_\bot~  [
  \frac{1}{2}\left( \frac{\partial \phi}{ \partial \tilde{t}}\right)^2
 + \frac{1}{2}
 \left(\frac{\partial \phi}{ \partial \tilde{r}} \right)^2+\frac{1}{2}
 \left( {\nabla_\bot \phi}\right)^2
 +  V(\phi)]\} \no
 &=&N_1\int
  D\phi ~exp\left\{-  \frac{I_E}{\hbar}\right\}.
 \eeqa

where $I_E$ is the Euclidean action for the scalar field in the inertial frame.
Since both of $Z_R$ and $Z_Q$ can be obtained from $Z^E_Q$ by analytic continuation, they are
physically equivalent ~\cite{Crispino:2007eb}.
Thus, the conventional
  QFT formalism is equivalent to the purely  information theoretic formalism
  for loss of information about field configurations beyond the Rindler horizon.

Recall that Eq. (\ref{Z_R}) was derived without using any quantum physics.
Since quantum mechanics can be thought of as the single particle limit of QFT,
this implies
that quantum mechanics emerges from the application of
information theory to Rindler horizons and
is not fundamental in our formalism.

Near any static horizon having more generic static metrics
$ds^2=-f^2 dt^2+ \gamma_{\alpha\beta} dx^\alpha dx^\beta$,
the metric reduces
to the Rindler form~\cite{Padmanabhan:2003gd}
$ds^2\simeq -f^2 dt^2 + df^2/\kappa^2 + dL_\bot^2$,
where $\kappa$ is the surface gravity and $dL_\bot^2$ is the metric for the orthogonal direction.
Therefore, we expect the information theoretic interpretation of QFT
to be valid for more generic static metrics.

This information theoretic approach is more than a reinterpretation of quantum mechanics.
For example, it could explain the origin of quantum randomness and PI quantization,
which were assumptions in ordinary quantum mechanics.
Note that by extremizing  $Z_R$
this approach also explains the origin  of the  thermodynamic relation
$dE=TdS$ in gravitational systems~\cite{myreview}, which leads to
entropic gravity and Jacobson's thermodynamic model of Einstein's gravity.
Another bonus is the explanation of the well-known  analogy between  QFT
and classical statistical mechanics; QFT is essentially a statistical system in disguise.
Surprisingly, this formalism also gives rise to a derivation of the holographic principle,
which will be presented in the next section.

\section{Holographic principle from information loss}

The information theoretic derivation of quantum mechanics  in the previous section
makes it simple
to understand the physical origin of  the holographic principle. Consider a $d+1$-dimensional bulk region $\Omega$ with a  $d$-dimensional boundary $\partial \Omega$ that
is a one-way
causal horizon (see Fig. 2) such as a black hole horizon, Rindler horizon or cosmic horizon.
Imagine an outside observer $\Theta_O$ (like $\Theta_R$) who can not access
information about matter or spacetime in the region
 because of the horizon.
  The situation in $\Omega$ is maximally uncertain to the outside observer,
   and the best  the observer could do
is to estimate the probability of each possible field configuration
 of $\phi$ in $\Omega$, which turns out to be the probability amplitude in the PI~\cite{myquantum}.
During this estimation, $\Theta_O$ would use  the
maximal information available to her/him.
The previous section showed
 that the observer's ignorance leads to quantum fluctuations of fields in $\Omega$.
Thus, paradoxically, the outside observer's ignorance is an essential ingredient for any physics in $\Omega$.

\begin{figure}[tpbh]
\includegraphics[width=0.4\textwidth]{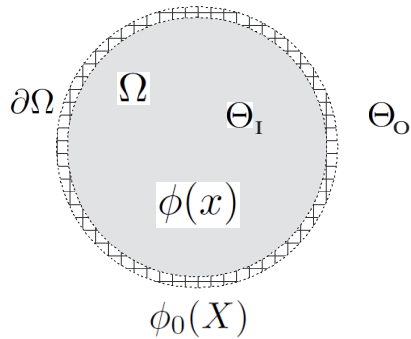}
\caption{Consider a bulk $\Omega$ with a causal horizon $\partial \Omega$ and an inside observer $\Theta_I$.
The outside observer $\Theta_O$ has no   information about the field configuration $\phi(x)$ in $\Omega$
except for its boundary values $\phi_0(X)$ and derivatives. Thus, according to the information theoretic interpretation,
 the physics in $\Omega$ is completely described
by the boundary physics on $\partial \Omega$, which is just the holographic principle.}
\end{figure}

According to the postulate 2, there is no non-local  interaction that might allow  super-luminal communication.
Therefore, we restrict ourselves to  local field theory in this paper.
For a local field, any influence on $\Omega$ from the outside of the horizon
should pass the horizon.
 This means that, according to  postulate 3,  all the physics in the bulk $\Omega$ is fully described by
 the DOF on the boundary $\partial \Omega$,
 which is just the essence of the holographic principle!
 In other words, information loss due to a horizon gives rise to quantum randomness in the bulk,
 and at the same time allows the outside observer  $\Theta_O$ to describe the physics in the bulk
 using only the DOF on the boundary. That is the best  $\Theta_O$ can do by any means, and the general
 equivalence principle demands that this description is sufficient  for understanding
 the physics in the bulk,
 which is  the holographic principle.
 %The DOF on the boundary contains the boundary value of $\phi$
% and some dual boundary field $\lambda$ having an action $S_\lambda(\lambda)$ corresponding to $H(\phi)$.

 % In short, redundancy exists in the DOF in $\Omega$.
Therefore, the following version of  the holographic principle is a natural consequence of the information theoretic formalism of QFT based on the  three postulates.
\begin{theorem}[holographic principle]
 For local field theory, physics inside a causal horizon can be  described
 completely by physics on the horizon.
\end{theorem}

One may think of this as  a $derivation$, albeit simple one, of the holographic principle from
the  information theoretic
postulates. Note that this derivation is  generic, because the arguments we used in this section  rely
on neither the specific form of the metric nor any symmetries the fields may have.

 What else can the  information theoretic  formalism tell us  about the holographic principle?
 First, the holographic principle cannot be applied to a general surface that is not a causal horizon.
 The range of application of the principle was a long standing problem.
Second, according to postulate 2, there should be a finite length scale $l$,
 and a finite area $l^2$ that can contain a bit of information.
Thus, the total entropy $S$ that the surface $\partial \Omega$, and hence  the bulk $\Omega$, can have is proportional to the surface area $A$;
\beq
S= \frac{A}{l^2}.
\eeq
The area law naturally emerges too.
Third, it implies that the black hole entropy represents the uncertainty of  field configurations inside the
black hole horizon.

How can we relate bulk physics with boundary physics?
 To demonstrate the plausibility of  theorem 1, I derive Witten's prescription of the holographic principle
 as an example.
 (Assume that  the scalar field $\phi$ has a Lagrangian
$L_0=\sqrt{-g}(\nabla^\mu\phi \nabla_\nu \phi-V(\phi))$.)
The conjecture says
\beq
\label{witten}
Z_{\partial \Omega}[\phi_0]=Z_{\Omega} [\phi|_{\partial \Omega} =\phi_0],
\eeq
where $Z_{\partial \Omega}[\phi_0]\equiv\langle exp(-\int d^d X \phi_0 \lambda)\rangle_{\partial \Omega}$
is the generating functional on $\partial \Omega$
with $\phi_0$ as a source for a boundary field $\lambda$.
$Z_{\Omega} [\phi|_{\partial \Omega} =\phi_0]$
 is the partition function for the bulk field $\phi$ on $\Omega$ which approaches
$\phi_0$ at the boundary.

In  Sec. II, we used only one nontrivial constrain on $E$.
More generally, other constraints may exist regarding boundary field values $\phi=\phi_0$ on $\partial \Omega$, which the outside observer
$\Theta_O$ can measure in principle. (When it is impossible to assign a field value on $\partial \Omega$, one may consider a stretched horizon instead of the horizon itself.)
Thus,
the field value $\phi_0$ and its derivatives at each point
 on the boundary $\partial \Omega$ could be the maximal information  (besides $E$ and the form of $H(\phi)$) that
 the observer $\Theta_O$ can measure or change to influence the physics in $\Omega$ and that constrains
 the probabilities for the field in $\Omega$.
 For simplicity, we assume a  Dirichlet boundary condition.
%Let  $Z_{\Omega} [\phi|_{\partial \Omega} =\phi_0]$ be the bulk partition function with this constraint.

Alternatively, one can describe the bulk physics by using only the quantities defined on the boundary.
Imagine that there is only one
 boundary field $\lambda$ that has an action $S_\lambda$ and an interaction term $ \phi_0 \lambda$.
  Then, the partition function
for the boundary field is
\beq
\label{Zlambda0}
Z_\lambda[\phi_0] = \frac{\int D\lambda ~e^{- S_\lambda} e^{-\int_{\partial \Omega} \phi_0 \lambda}}
{\int D\lambda ~e^{- S_\lambda} },
\eeq
which is  just  $Z_{\partial \Omega}[\phi_0]$. We have set $Z_\lambda[\phi_0=0]=1$.

 The effective number of DOF of $\lambda$ should be equal to that of $\phi$, because
theorem 1 implies that the boundary physics with
 $(S_\lambda(\lambda),E,\phi_0)$ has  all the information about the bulk field having $(E,\phi_0,H(\phi))$ as parameters.
For a given $E$ and $H$, the classical field $\phi_0$ is the only free parameter describing both of the bulk partition function
and the boundary partition function.
 The partition function for a thermal system
should contain all the information of the system. Since $\lambda$ should describe the physics of $\phi$,
$Z_{\lambda}[\phi_0]=Z_{\Omega} [\phi|_{\partial \Omega} =\phi_0]$
should hold for $\lambda$ having $\phi_0$ as a source. Here, $S_\lambda$ is not arbitrary
but should be such  that $Z_{\lambda}[\phi_0]$ well reproduces
$Z_{\Omega} [\phi|_{\partial \Omega} =\phi_0]$.
 In other words, there is a duality mapping $(\phi_0,H(\phi))\longleftrightarrow(\phi_0,S_\lambda(\lambda))$.
$\lambda$ could be  hypothetical (mathematical) rather physical.
In short, Witten's prescription is a natural consequence of the information theoretic formalism.
However, the derivation above does not guarantee the existence of ($\lambda,S_\lambda$) for arbitrary bulk fields.
Since a description is not a physical rule, the derivation is enough for our purpose.

 Of course, the  saddle point approximation of the bulk
partition function becomes
\beq
Z_\lambda[\phi_0]\simeq exp(-I_E(\phi)),
\eeq
where $I_E(\phi)$ is the Euclidean classical action 
with the boundary condition $\phi|_{\partial \Omega}=\phi_0$ in the curved spacetime,
as usual. Then, Witten's prescription yields  a relationship between QFT and gravity.

To be concrete, let us consider a derivation of the prescription for the Rindler metric in detail as an example.
 To show the equivalence   we divide the surface $\partial \Omega$ into $N_j$ small patches
 and  discretize   the bulk field  $\phi$.
 We also assume that the field satisfies a Dirichlet boundary condition at the horizon.
By repeating the calculation leading to Eq. (\ref{P}) with additional constraints on the
expectation values $\sigma_{j}$ of the boundary field
$\phi_0$ at the $j$-th patch $\phi_{0j}$,
\beq
\langle \phi_{0j} \rangle \equiv \sum_i P[\phi_i] \phi_{0j}(\phi_i)=\sigma_{j},~(j=1\cdots N_j),
\eeq
 one can easily obtain
 the probability distribution
 \beq
 P[\phi_i]=\frac{1}{Z} exp\left[-\beta H (\phi_i)-\sum_{j=1}^{N_j} \lambda_j \phi_{0j}(\phi_i)\right],
 \eeq
 where  $\phi_{0j}(\phi_i)$ represents a boundary field value at the $j$-th patch corresponding to
 a specific bulk field configuration $\phi_i$, and $\lambda_j$ is the Lagrange multiplier field
 at  patch $j$. Of course
   $Z=\sum_{i=1}^n  \exp\left[- {\beta H(\phi_i)}-\sum_{j=1}^{N_j} \lambda_j \phi_{0j}(\phi_i) \right]$ now.
   The index $j$ denotes the position on $\partial \Omega$ and shall be promoted to the $d$-dimensional
   coordinate $X$ in the continuum limit.
   Since the number of independent $\lambda_j$ values is $N_j$ and  $\lambda_j$ couples to the boundary field,
   we can naturally think of $\lambda_j$ as another scalar field  on the boundary $\partial \Omega$.
 Taking a continuum limit and repeating the procedure leading to  Eq. (\ref{ZEQ})
 we obtain
 \beqa
\label{ZEQ2}
 Z^E_Q [\sigma]&=&  N_1\int
  D\phi(x) ~exp\{- \int d\tilde{r}d\tilde{t} dx_\bot~  [L_0(\phi)+\lambda(X) \sigma(X)]\},
 \eeqa
where $L_0$ is the Euclidean Lagrangian and
 $\lambda(X)$ and $\sigma(X)$ are promoted continuous versions of  $\lambda_j$ and $\sigma_{j}$,
respectively.

Note that this term is just the right-hand side of  Eq. (\ref{witten}), $Z_{\Omega} [\phi|_{\partial \Omega} =\sigma]$
with $\langle \phi_0(X) \rangle=\sigma(X)$ identified
 as a classical boundary value for $\phi$ at $\partial \Omega$.
This identification is physically reasonable because, strictly speaking,  our theory and  ordinary QFT
 contain  no genuine classical field.
The classical field is an approximate concept valid only in a specific limit.

Next, we need to show that $Z^E_Q[\sigma]$ is equivalent to $Z_{\partial \Omega}[\sigma]$.
Since $\sigma(X)$ is a c-function defined on the boundary, it has no-dynamics on $\partial \Omega$.
Thus, one can think of  $\sigma(X)$ as a source function linked with some boundary field.
Considering the action in $Z^E_Q$, the simplest candidate for the boundary field
 is the Lagrange multiplier field $\lambda(X)$ itself.
This dual field may have an action $S_\lambda(\lambda)$ on $\partial \Omega$. Therefore, the partition function
for the boundary field is
\beq
\label{Zlambda}
Z_\lambda[\sigma] = \frac{\int D\lambda ~e^{- S_\lambda} e^{-\int_{\partial \Omega} \sigma \lambda}}
{\int D\lambda ~e^{- S_\lambda} },
\eeq
which is  $Z_{\partial \Omega}[\phi_0]$ .

Within the conventional QFT formalism, it seems impossible to prove the general equivalence of $Z_\lambda[\sigma]$
and $Z^E_Q[\sigma]$ because of the difference in their spacetime  dimensions.
Here, we  must recall theorem 1.
According to the theorem, all the information in the bulk
should be contained in the boundary, and $\sigma$  should
be the only messenger available (except for $E$) between the bulk and the boundary.
(We also need to link $S_\lambda$ to $H$.)
In other words, regarding $\phi$ in $\Omega$ and $\partial \Omega$, the outside observer can change only $\sigma$.
Furthermore, the  relevant fields describing  the effectively  same system  (the bulk and boundary scalars) are $\phi$ and $\lambda$.
Thus, the two partition functions  as  functionals of $\sigma$ should be equal, i.e.,
$Z_\lambda[\sigma]=Z^E_Q[\sigma]$ and Witten's prescription Eq. (\ref{witten}) holds for the Rindler metric.
The generating functional and the Euclidean nature of Witten's prescription
arise naturally in the formalism.

\section{Conclusions}

In summary,
this paper shows that the holographic principle, like  quantum mechanics and gravity,
is not fundamental but emerges from information loss
at causal horizons.
The derivation is generic because we assumed neither supersymmetry nor string theory.
This suggests the universality of the holographic principle applied at causal horizons
and validates the application of the principle to other quantum systems such as condensed matter.
The principle is intimately related to quantum mechanics.
The derivation of the holographic principle in this paper
 is not a simple transformation of the principle to the postulates,
because with the postulates one can derive quantum mechanics and Einstein's gravity as well as the principle.
Since quantum mechanics and holography
originate  in information loss at causal horizons,
 information seems to be the common  root of physics.
This could open a new route to unifying gravity and quantum mechanics.

In our future work, we need to verify the equivalence of the information theoretic formalism
and QFT in a more generic curved spacetime.
To check the usefulness of this formalism,
it is desirable to show the relationships between partition functions for other spacetimes, especially
 the AdS/CFT correspondence.

\section*{acknowledgments}
This work was supported in part by Basic Science Research Program through the
National Research Foundation of Korea (NRF) funded by the ministry of Education, Science and Technology
(2010-0024761) and the topical research program (2010-T-1) of Asia Pacific Center for Theoretical
Physics.
%
%\bibliographystyle{h-physrev}
%\bibliography{entanglement}

\end{document}